\documentclass[a4paper]{jpconf}
\usepackage{graphicx}
\usepackage{amsmath}
\usepackage{amssymb}
\usepackage{epsfig}
\graphicspath {{./figures/}}

\begin{document}
\title{Efficient algorithms for computing ground states of the 2D random-field Ising model}
\author{Argyro Mainou$^1$, Nikolaos G Fytas$^1$ and Martin Weigel$^{2,1}$}
\address{$^1$ Centre for Fluid and Complex Systems, Coventry
	University, Coventry, CV1 5FB, United Kingdom}
\address{$^2$ Institut für Physik, Technische Universität Chemnitz, 09107 Chemnitz, Germany}
\ead{mainoua@uni.coventry.ac.uk}

\begin{abstract}
We investigate the application of graph-cut methods for the study of the critical behaviour of the two-dimensional random-field Ising model. We focus on exact ground-state calculations, crossing the phase boundary of the model at zero temperature and varying the disorder strength. For this purpose we employ two different minimum-cut--maximum-flow algorithms, one of augmenting-path and another of push-relabel style. We implement these approaches for the square and triangular lattice problems and compare their computational efficiency.
\end{abstract}


\section{Introduction}
\label{sec:intro}
The random-field Ising model (RFIM) is one of the simplest disordered systems~\cite{Imry_1975, Aharony_1976, Young_1977, Parisi_1979}. Its study not only allows the investigation of a series of complex phenomena, but also is directly connected to a significant number of experiments in condensed-matter physics~\cite{Rieger_1995_2, Belanger_1998, Vink_2006}. These facts have established it as one of the main model systems for the study of collective behaviour under quenched disorder. It is described by the Hamiltonian:

\begin{equation}
\begin{split}
\mathcal{H} = -J\sum_{\langle x,y \rangle}s_{x}s_{y} - \sum_{x}h_{x}s_{x},
\end{split}
\label{eq:three_one}
\end{equation}

\noindent where $s_{x} = \pm 1$ and $J>0$ denotes the ferromagnetic nearest-neighbour interaction (we set $J=1$ throughout this study). The disorder is introduced via the variables $h_x$, which are independent random magnetic fields acting on each spin $s_x$ and following some distribution $\mathcal{P}(h_x)$. In the present work, we consider quenched random fields following the Gaussian distribution ($\mu =0$, ${\sigma}^2 =h^2$), so that $h$ defines the disorder strength. 

\begin{figure}[tb!]\vspace{-0.7cm}
    \centering
	\includegraphics[width=90mm]{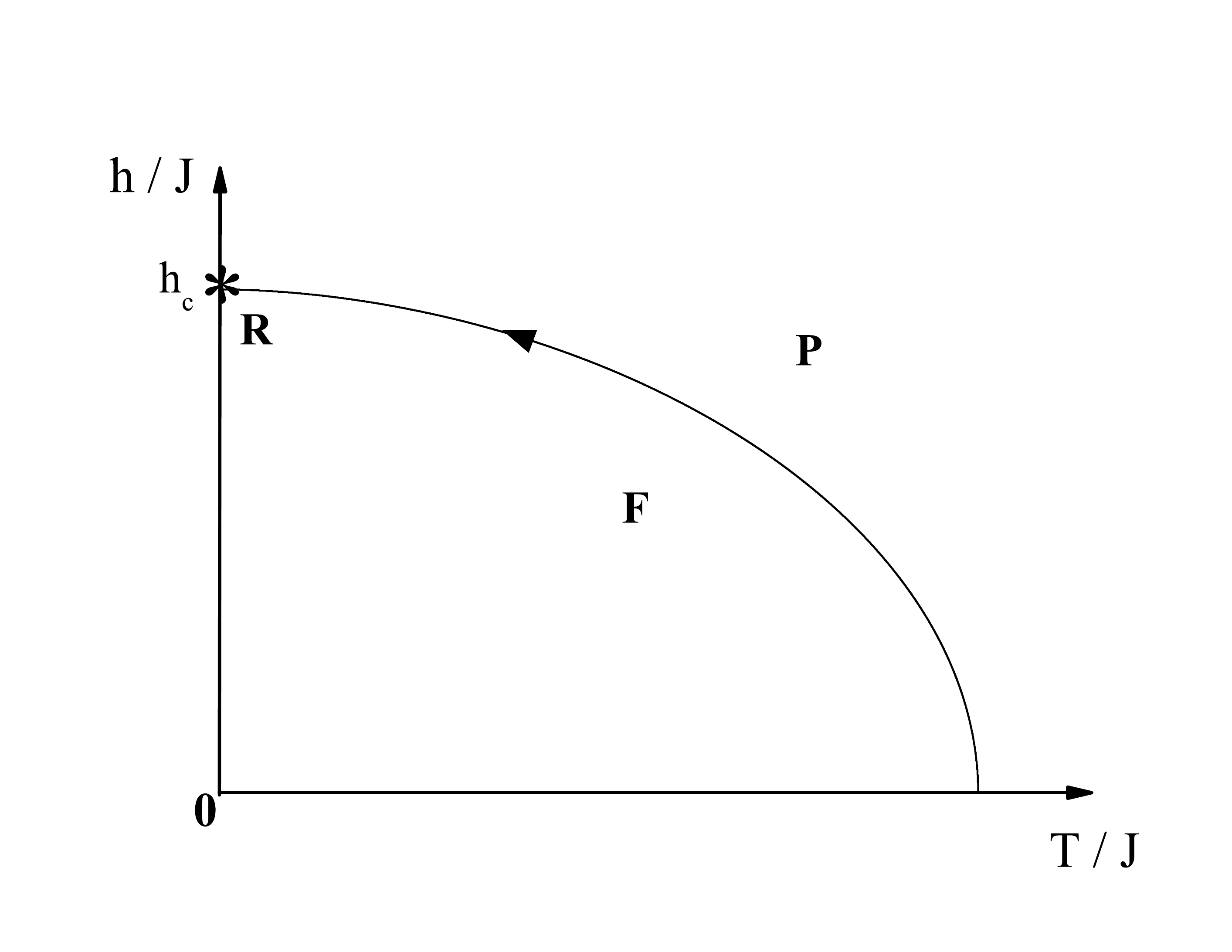}
	\vspace{-0.4cm}
	\caption{A sketch of the phase diagram of the RFIM. The solid line (phase boundary) separates the ferromagnetic (\textbf{F}) and paramagnetic (\textbf{P}) phases. The black arrow shows the renormalization-group flow to the random fixed point (\textbf{R}) at $T=0$ and $h = h_{\rm c}$, the critical disorder strength, as marked by an asterisk.}  
    \label{fig:phase_diagram}
\end{figure}

A number of numerical techniques, mainly of Monte-Carlo type, have been employed for studying the critical (and other) properties of the RFIM~\cite{Rieger_1995_2}. However, the limitations of these methods due to the rough free-energy landscape of the model led the community to seek other more efficient methods. Fortunately, the renormalization group tells us that the random field is a relevant perturbation at the pure fixed point, and the random-field fixed point is at $T = 0$ (see figure~\ref{fig:phase_diagram})~\cite{bray_1985}. Hence the critical behavior is the same everywhere along the phase boundary and we can predict it simply by staying at $T = 0$ and crossing the phase boundary at the critical field point. This is a convenient approach because we can determine ground states of the system exactly using efficient optimization algorithms through an existing mapping of the ground-state problem to a maximum-flow optimization task~\cite{Rieger_1998,Cormen_1990,Papadimitriou_1994}.

In 1975, Picard and Ratliff introduced the idea of treating energy minimization problems of a class of systems, including the RFIM, as minimum-cut problems on a network~\cite{Picard_1975}, by mapping the system under consideration onto a network and calculating its maximum flow. The value of the maximum flow is then equal to the ground-state energy of the system. The orientations of the spins in the ground state can be constructed from the flow values of the corresponding network. These techniques have allowed the extensive employment of network theory for calculating ground states~\cite{Swift_1997, Seppala_2001_2, Stevenson_2011, Fytas_2013_2}.

Motivated by the aforementioned results, we apply graph-cut theory in order to address such energy minimization problems for the Gaussian RFIM in two dimensions. For this purpose, we employ two distinct minimum-cut-maximum-flow algorithms. The most popular algorithms of this type are roughly divided into two main categories: augmenting-path-style and push-relabel-style. We chose an algorithm from each category and implement it for the study of the RFIM, as outlined in  section~\ref{sec:methods}. The applicability of the augmenting-path-style algorithm is examined for two lattice geometries of the model, namely square and triangular and the results are presented in section~\ref{sec:results}, where we also perform a comparison between the two algorithms in terms of their computational time for the square-lattice model. Finally, a summary of our results is provided in  section~\ref{sec:summary}. 


\section{Numerical Methods}
\label{sec:methods}

In this section we provide an outline of our numerical implementation on various random-field systems with periodic boundary conditions. As a starting point, we point out that the mapping of the model to a network was carried out according to reference~\cite{Hartmann_2004} and that we used the Mersenne Twister~\cite{Matsumoto_1998} random-number generator (RNG) for the production of the Gaussian local random-field values.

Let us start with the Boykov-Kolmogorov (BK) algorithm~\cite{Boykov_2004}, which is a variation of standard augmenting-path algorithms~\cite{Ford_1956, Dinic_1970, Edmonds_1972} and was developed in order to improve their empirical performance on graphs in computer vision problems. Its input is a weighted, directed, residual graph, along with a source $s$ and a sink $t$. BK involves the construction of two non-overlapping search trees, that is a source tree and a sink tree, which treats the terminals symmetrically and so outperforms an earlier version of the algorithm where a single tree rooted at the source was used~\cite{Boykov_2001}. What is more, instead of the usual process of building new augmenting paths at each new iteration, the above trees are reused. We refer the interested reader to reference~\cite{Boykov_2004} for further details regarding the algorithm. The worst case complexity of BK is $\mathcal{O}(mn^2|C|)$, where $n$ is the number of edges of the network, $m$ is the number of nodes and $|C|$ is the cost of a minimum cut, which is the value of the maximum flow. Boykov and Kolmogorov showed~\cite{Boykov_2004} with experimental tests in two dimensions and for lattice sizes of up to $L=511$, that despite the fact that theoretically speaking this is worse than the time complexities of other standard minimum-cut--maximum-flow algorithms~\cite{Ford_1956, Dinic_1970, Edmonds_1972, Goldberg_1988, Cherkassky_1997}, the BK algorithm is significantly faster when applied to typical system sizes in real applications.

In  the  present  work we validate their results for the case of the square-lattice RFIM. Before doing so, we examine the algorithm's efficiency in working out the model's ground-state energy and spin configuration, not only for the square but also for the triangular geometry. In order to achieve this goal, we adapt the code for BK  provided in~\cite{Boykov_2004} (version 3.01) to the problem of computing RFIM ground states for square and triangular lattices. We pick $h=1$ for the square lattice and $h=1.7$ for the triangular lattice as test cases, since these values correspond to the same breakup length $\ell_b \approx 32$  for the two lattice geometries \cite{Binder_1983}.

\begin{figure}[tb!]
    \centering
    \includegraphics[width=39mm]{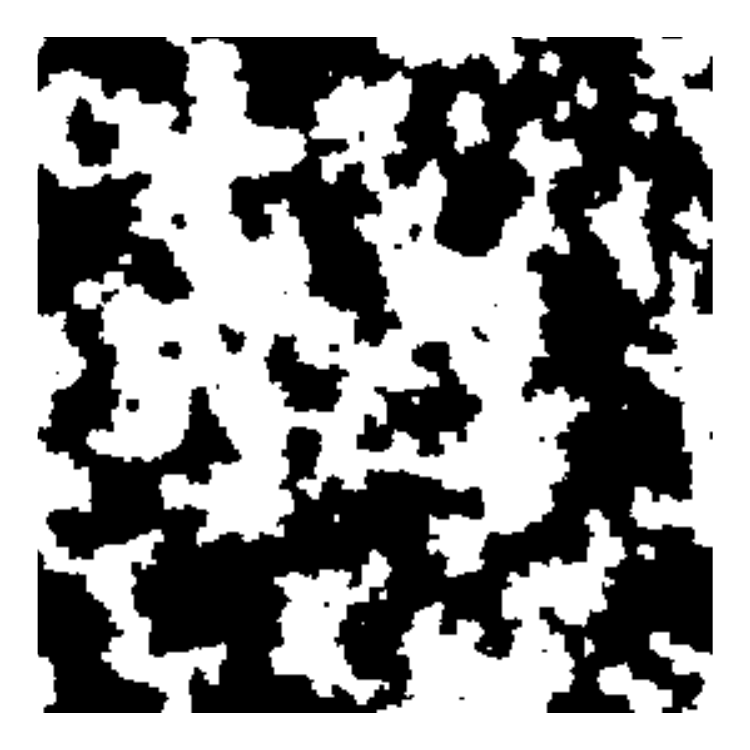}
	\includegraphics[width=39mm]{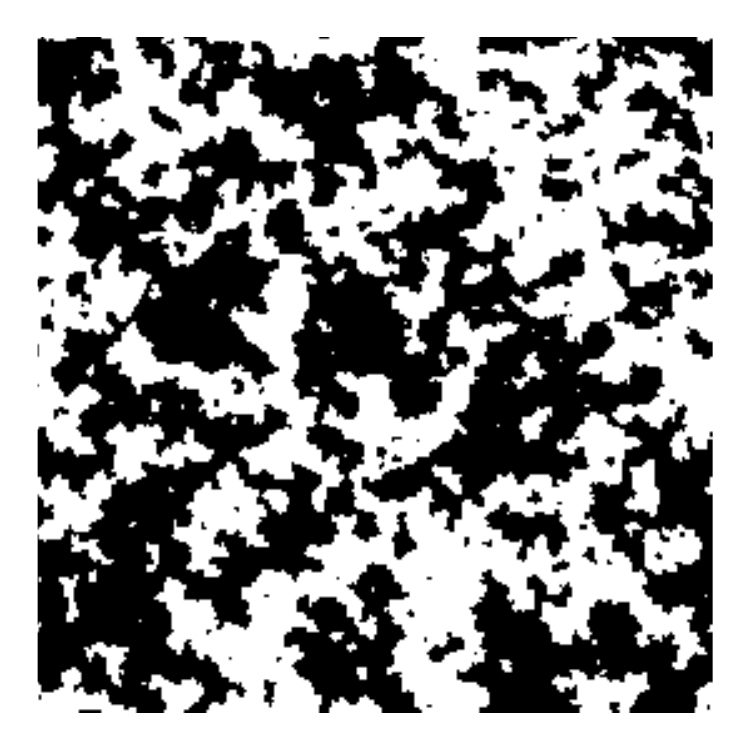}
	\includegraphics[width=39mm]{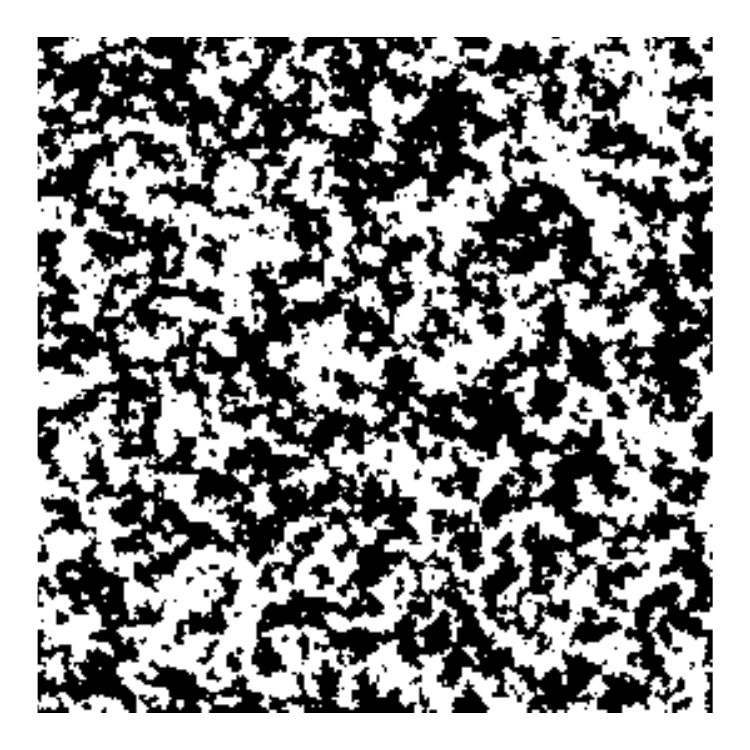}
	\includegraphics[width=39mm]{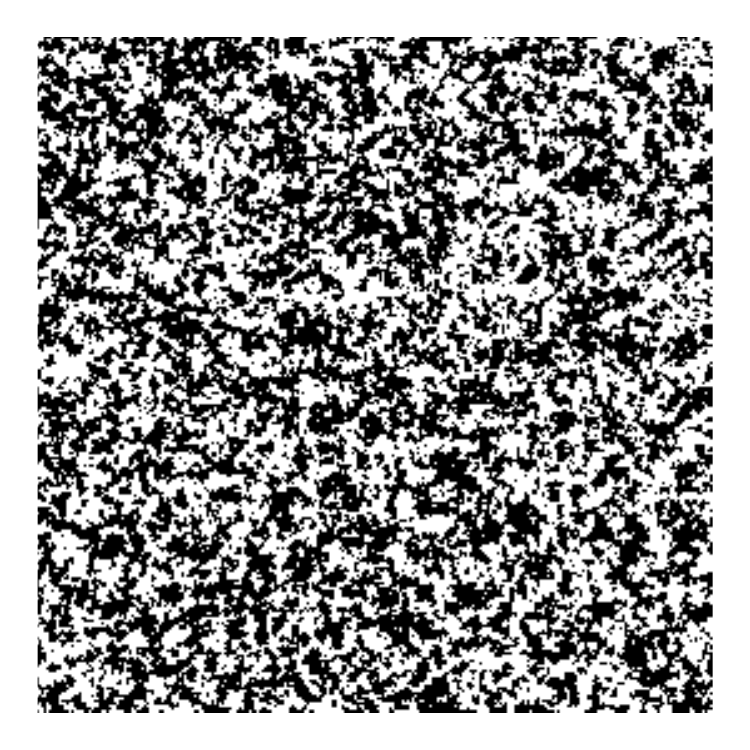}\\
	\vspace{0.5cm}
	\includegraphics[width=160mm]{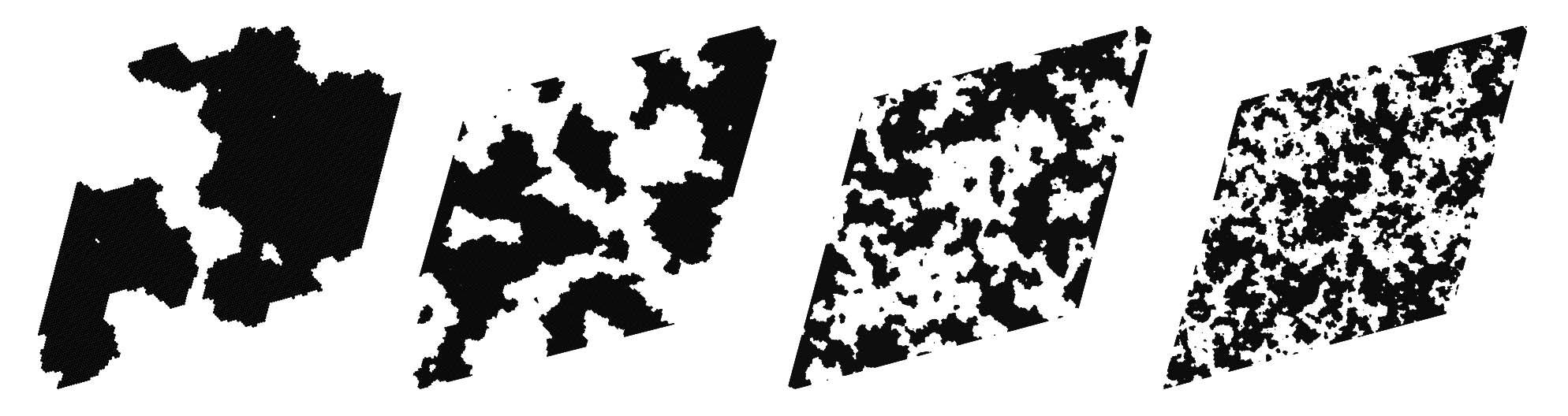}
	\caption{\label{fig:GS_spins} Example ground-state spin configurations of the RFIM on the square lattice (top row, $h=1$) and the triangular lattice (bottom row, $h=1.7$). System sizes are $L=512$, $1024$, $2048$ and $4096$ (top row), and $L=128$, $256$, $512$ and $1024$ (bottom row).}
\end{figure}

We now turn to the second category of minimum-cut-maximum-flow algorithms, namely algorithms of push-relabel-style. One of the most popular network flow algorithms used to solve the RFIM is the general push-relabel algorithm~\cite{Goldberg_1988} with a computational time complexity of $\mathcal{O}(n^2m)$. Proofs and theorems regarding the algorithm have been provided in references~\cite{Cormen_1990, Papadimitriou_1994}. Two implementations of the push-relabel method are the H\_PRF and Q\_PRF techniques~\cite{Cherkassky_1997}, which outrun many other minimum-cut-maximum-flow algorithms in real-size experiments, with computational time complexities of $\mathcal{O}(n^2\sqrt{m})$ and $\mathcal{O}(n^3)$, respectively. However, according to~\cite{Boykov_2004}, this does not hold for the case of the BK algorithm. With the aim of confirming this, here we choose to focus on a Q\_PRF-style modification~\cite{Middleton_2001, Middleton_2002} that removes the source and sink nodes, reducing memory usage and clarifying the physical connection. In particular, we follow the implementation described in reference~\cite{Fytas_2016}.

\begin{figure}[tb!]
    \centering
	\includegraphics[width=79mm]{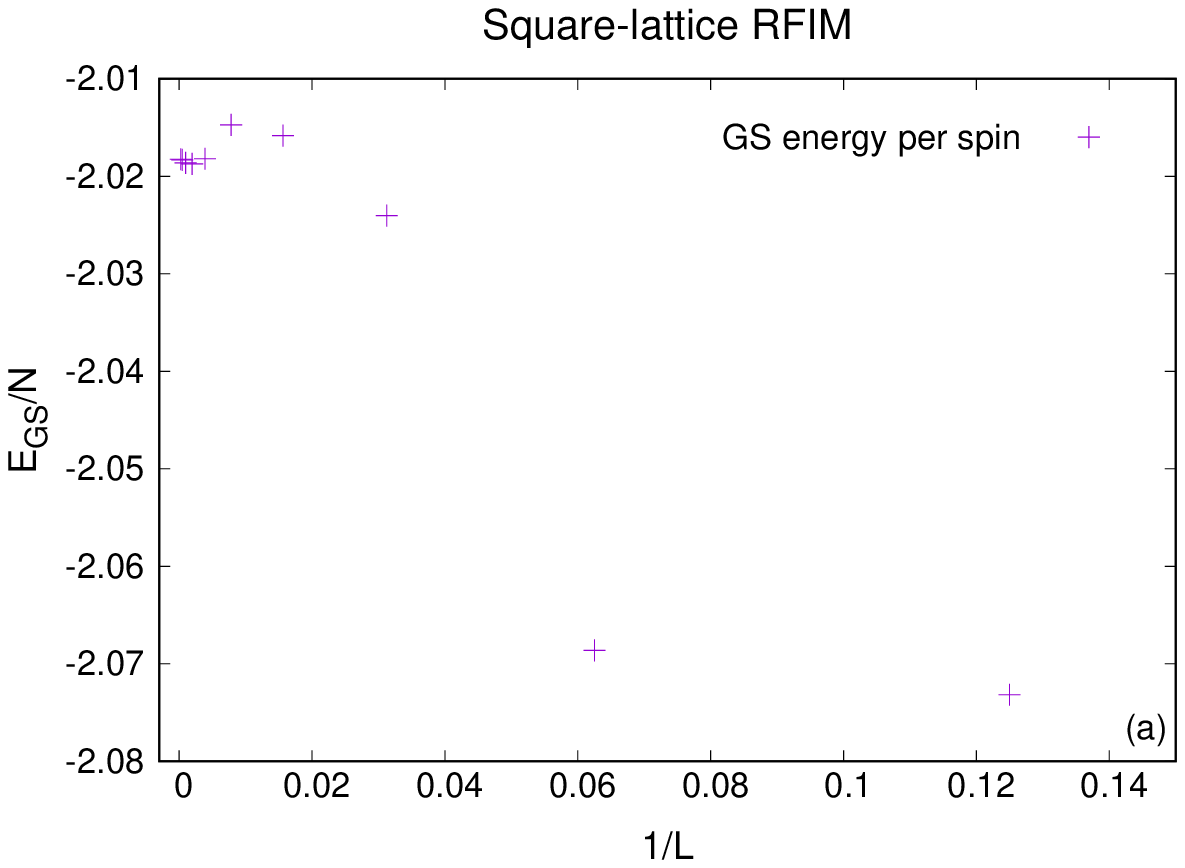}
	\includegraphics[width=79mm]{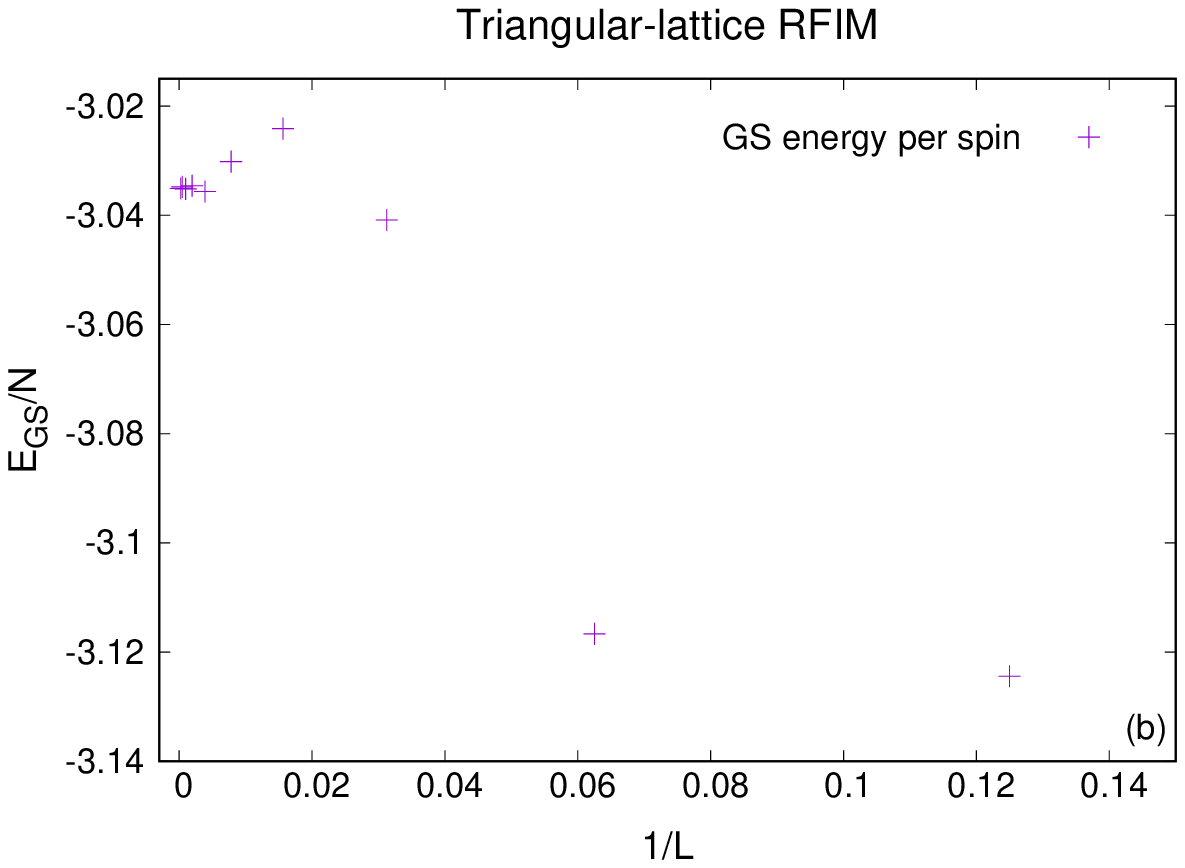}
	\caption{\label{fig:GS_energies} Ground-state (GS) energy per spin, $E_{\text{GS}}/N$ (where $N=L\times L$), of single realizations of the RFIM (a) on the square lattice and (b) on the triangular lattice for linear system sizes $L \in \left\lbrace 8,16,32,64,128,256,512,1024,2048,4096 \right\rbrace$, where $h=1$ and $h=1.7$ for the square and the triangular lattices, respectively.}
\end{figure}

The implementation of the BK algorithm for creating multiple realizations of the quenched disorder can be carried out in two different ways, that is by either building the code's graph from scratch for every single disorder sample or building the graph once and reallocating the graph's edge capacities for every disorder realization. We expect that the latter constitutes an optimized version of the code, since the graph needs to be constructed outside of the loop for the many samples of the disorder. We therefore examine both versions and our results are provided in subsection~\ref{sec:results_b}, along with the ones referring to the Q\_PRF-style algorithm.  Note that in order to ensure that the timing measurements are not perturbed by concurrent jobs, we reserve a full node on the cluster while only using a single core at each point in time.
\section{Results }
\label{sec:results}
\subsection{Boykov-Kolmogorov algorithmic efficiency: square- and triangular-lattice RFIM}
\label{sec:results_a}


We examine the BK algorithm's suitability for the study of the ground-state problem of the two-dimensional RFIM with the use of graph-cut methods. For this purpose, we implement it in order to work out the maximum flow of a network corresponding to a Gaussian $L \times L$ RFIM, for various lattice sizes $L$, of either square- or triangular-lattice geometry. In figure~\ref{fig:GS_spins} we show the spin configurations for some example disorder realizations on square lattices of linear lattice sizes $L=512$, $1024$, $2048$ and $4096$ and random-field strength $h=1.0$ as well as triangular lattices of sizes $L=128$, $256$, $512$ and $1024$ and random-field strength $h=1.7$, respectively. Note that for a given lattice size $L$, a larger disorder strength is required for the triangular case such that the sizes of the spin clusters are comparable between the two different geometries. For example, for $L=1024$, the square- and triangular-lattice RFIMs exhibit similar behaviour for $h=1$ and $h=1.7$, respectively. This is expected, since the two extra bonds between the nearest neighbours of the triangular lattice contribute to the resistance of breaking the ground-state order. From the given spin configurations, it is possible to extract the ground-state energies, which we present (also for single disorder realizations) in figure~\ref{fig:GS_energies}.

\begin{figure}[tb!]
    \centering
	\includegraphics[width=0.65\linewidth]{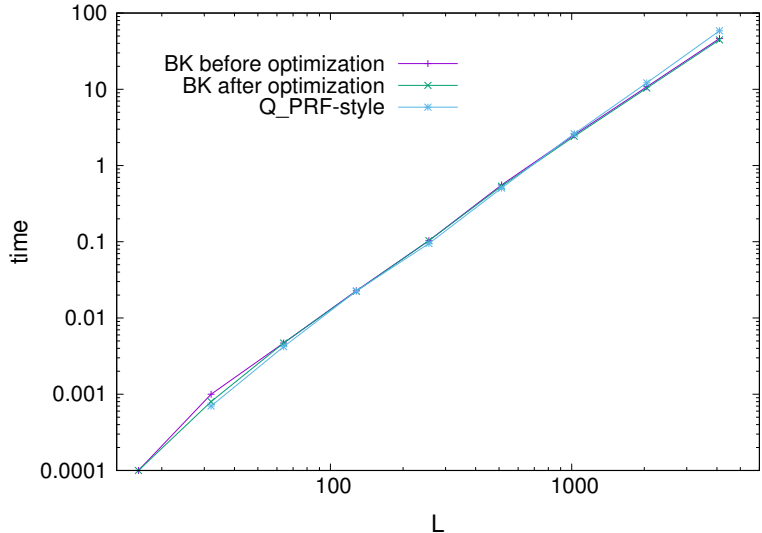}
	\caption{Average run-time in seconds for computing a single square-lattice ground state using the Q\_PRF-style algorithm of~\cite{Fytas_2016} (blue line) as well as  the BK algorithm before (violet line) and after (green line) optimization. The data are averaged over 100 disorder samples for $h=1$.}  
    \label{fig:BK_computational_time_complexity}
\end{figure} 

\subsection{Algorithmic comparison: square-lattice RFIM}
\label{sec:results_b}

We proceed to the inspection of both algorithms' time complexity for the square-lattice RFIM, for several lattice sizes of up to $L=4096$. For this purpose we generate many realizations of the quenched disorder (exactly the same for the two algorithms) for some specific disorder strength (we pick $h=1$) and work out the computational time of the codes for treating each one of them. Afterwards, the calculation of the mean value of those computational times will give us an estimate of the average computational time needed by each algorithm for working out a single sample's ground-state energy and corresponding spin configuration for $h=1$.

In figure~\ref{fig:BK_computational_time_complexity} we show the results regarding the computational time of the two versions of the BK algorithm compared to the ones we obtain by implementing the Q\_PRF-style algorithm, using the same disorder samples.
As expected, we find that the approach of  building the graph just once outruns the method of building the graph for every realization of the disorder and therefore it may be preferred for large-scale simulations as an optimized version of the code. It is clear however, that both versions of the BK algorithm are somewhat faster than the Q\_PRF-style algorithm. Performing fits of a power-law $t \sim L^a$ to the run-time data, we find $a\approx 2.3$ for all three codes. Note that this is a considerably slower increase than the worst-case scaling discussed above. Finally, one may notice that the difference between the two versions of the BK code implementation is relatively small. This is because the biggest portion of the computational time is used for the calculation of the maximum flow and not for setting up the graph structure.


\section{Summary and Outlook}
\label{sec:summary}

Motivated by the continuing use of graph-cut methods for the investigation of ground-state properties of the random-field magnets, we performed a comparative test of two different minimum-cut--maximum-flow algorithms on the Gaussian two-dimensional random-field Ising model. We implemented the particularly efficient Boykov-Kolmogorov algorithm \cite{Boykov_2004} and applied it to the square-lattice and triangular-latice random-field Ising models. Additionally, an optimization of the code was developed and both versions were compared to a Q\_PRF-style implementation of the push-relabel algorithm, currently reported as one of the fastest minimum-cut-maximum-flow algorithms. We showed that the Boykov-Kolmogorov method is more efficient than the Q\_PRF approach when implemented for square lattice sizes up to $L = 4096$, a fact that verifies and extends the original claim of reference~\cite{Boykov_2004} that the Boykov-Kolmogorov algorithm outruns many of the most efficient minimum-cut-maximum-flow algorithms for lattice sizes of experimental interest.


\ack
We acknowledge the allocation of CPU time on the supercomputer \emph{Zeus} of Coventry University. A. Mainou would like to thank Abhishek Kumar for his support at various stages of this work.

	
	\section*{References}
	
	\bibliographystyle{iopart-num}
	
	\providecommand{\newblock}{}

\end{document}